\documentclass[a4paper,10pt,onecolumn,english]{article}
\usepackage{babel}
\usepackage[utf8]{inputenc}
\usepackage[T1]{fontenc}
\usepackage{hyperref}
\usepackage{tabularx}
\usepackage{graphicx}
\usepackage[ruled,vlined,linesnumbered]{algorithm2e}
\usepackage{bm}
\usepackage{todonotes}
\usepackage{amsmath}
\usepackage{amsthm}
\usepackage{cancel}
\usepackage{appendix}

\newcounter{protocol}
\newenvironment{protocol}[1]
  {\par\addvspace{\topsep}
   \noindent
   \tabularx{\linewidth}{@{} X @{}}
    \hline
    \refstepcounter{protocol}\textbf{Protocol \theprotocol} #1 \\
    \hline}
  { \\
    \hline
   \endtabularx
   \par\addvspace{\topsep}}

\newcommand{\sbline}{\\[.5\normalbaselineskip]}

\newtheorem{theorem}{Theorem}[section]

\title{Towards Better Privacy-preserving Electronic Voting System}

\author{
    Zipeng Yan, 
    Zichao Jiang,
    Yiyuan Li\\
	\footnotesize Computer Science Department, University of North Carolina, Chapel Hill
}

\date{\empty} 

\begin{document}
\maketitle
\begin{abstract} 
This paper presents two approaches of privacy-preserving voting system: Blind Signature-based Voting (BSV) and Homorphic Encryption Based Voting (HEV). BSV is simple, stable, and scalable, but requires additional anonymous property in the communication with the blockchain. HEV simultaneously protects voting privacy against traffic-analysis attacks, prevents cooperation interruption by malicious voters with a high probability, but the scalability is limited. We further apply sampling to mitigate the scalability problem in HEV and simulate the performance under different voting group size and number of samples.

\end{abstract}
\section{Introduction}

Voting is an essential element of the current social and political systems. Individual voting result is a personal privacy which voters may be unwilling to release publicly, especially to the supports of the lost candidate or proposal. The prevailing of electronic voting systems with tracable electronic traces increase the concern of privacy disclosure.

Currently there are many popular privacy-preserving techniques, such as multi-party computation and differential privacy. However, in real-world voting, the security of the former methods relies on splitting an authority into multi-parties, which usually lacks effective supervisions. Meanwhile, The decision-making of a voting process is threshold-based. Therefore, it is hard to deploy differential privacy-based methods, which requires a less sensitive information accuracy of individual votes.

In this paper, we present two approaches towards privacy-preserving electronic voting system. The first approach is a blind signature based voting (BSV), where we apply blind signature to ensure each vote only comes from eligible voters and the government is not able to access the content. The second approach is homomorphic encryption based voting (HEV), where we use homomorphic encryption to calculate the encrypted votes directly and hide individual votes. We also update HEV into homomorphic encryption voting with sampling (HEVS), which has resistance on malicious voters involved who try to ruin the result.

We explain the background and our threat models in Section~\ref{background}. In Section~\ref{Protocols} we briefly discuss our two protocols. We show possible loopholes for both of the approaches and possible solutions in Section~\ref{Improvement}. And we demonstrate and explain our experiment results in Section~\ref{Experiment Results}.

\section{Background}\label{background}

In our setting, voting consists of at least two parties: voters and the government. Voters are the main content of voting who submit ballots. The government is the authority determining the eligibility of voters, counting ballots, and publishing the results. We assume that government is able to identify all eligible voters and ignores the votes from any other ineligible voters.

Privacy-preserving voting aims at protecting voters' voting result(e.g. for or against) from any others, including the government. The government gets the overall result only, instead of individual results. 

Our voting environment is electronic, where voters and the government communicate through electronic devices\footnote{computers, databases, etc.}. During the communication, everything, including ballots, can be recorded. Thus, anyone who has recorded individual data can never revealed the result of its vote.

\subsection*{Threat Models}
Our solution assumes two levels of adversaries: semi-honest voters and semi-honest government will cooperate together to get information but do not deviate from the protocol specification; malicious voters and malicious government may arbitrarily deviate from the protocol execution and do whatever they want. 

Before we introduce our protocol, we argue that both of our protocols are privacy-preserving under two levels of adversaries. However, there are ways for malicious adversaries to halt the whole process. We provide combats in Section~\ref{Improvement}



\section{Two Privacy-preserving Approaches}\label{Protocols}


\subsection{Blind Signature Based Voting}
Blind signature is a form of digital signature where messages are signed without revealing the content to the signer~\cite{chaum1983blind}. The message is blinded using a secret blinding factor from the user prior to the signature. The signer generates a signature for the blinded message. Then the user uses the blinding factor to unblind the message and the signature, resulting in the original message with a legitimate signature. Throughout the process, the signer has no knowledge about the message without accessing the blinding factor. Blind signature is supported by many cryptographic algorithms including RSA and Elliptic Curve.

\subsubsection*{Protocol Blind Signature Voting (BSV)}
\textit{Notation.} \\
$\mathcal{V} = \{V_1, V_2, ... V_i, ... V_{n}\}$ is the group of eligible voters. $\forall i \in \{1, 2, ... n\}$, voter~$V_i$ holds a vote~$m_i$.\\
$Gt$ denoted the government who can identify these $n$ voters. \\
$B$ denoted a decentralized blockchain.\\
\textbf{Protocol detail in appendix~\ref{alg: blind sign based}}\\

In Protocol~\ref{alg: blind sign based}, every voter do a blind signature protocol with the government. Then voters publish their unblinded vote with the signature of the government to a blockchain through some anonymous channel. The blockchain is used $B$ to make the published ballots unmodifiable. After a period of time, $B$ stop collecting ballots and anyone can check the blockchain to know the voting result.
 
There are many projects of secure voting system using blind signature and blockchain~\cite{8726645}~\cite{liu2017voting}. 

\subsection{Homomorphic Encryption Based Voting}

Homomorphic encryption is a cryptographic tool of encryption model that allows computation on ciphertexts~\cite{gentry2009fully}. To achieve a privacy preserving electronic voting system, encryption of ballots is needed. And the ballot counting is an additive computation. Therefore, homomorphic encryption can be used to compute ciphertexts \footnote{the ballots} directly, with every ballot encrypted, which inherently protects voters' privacy. Moreover, additive homomorphic encryption is the only encryption involved.

In Protocol~\ref{alg: homo enc based}, we use a modified version of Elgamal encryption with threshold decryption~\cite{10.1007/3-540-39568-7_2}. Specifically, eligible voters generate secret key pieces $sK_i$ and use them to generate a public key $pK$ with the cooperation of the government $Gt$. Then voters use the public key to submit their vote, $0$ or $1$. The government aggregates the encrypted result and starts a threshold decryption to decrypt the aggregated result. During the entire process, the individual secret key $sK_i$ is not exposed to anyone. Therefore, individual result is safe and private in all circumstances.
\subsubsection*{Protocol Homomorphic Encryption Voting (HEV)}
\textit{Notation.} \\
$\mathcal{V} = \{V_1, V_2, ... V_i, ... V_{n}\}$ is the group of eligible voters. $\forall i \in \{1, 2, ... n\}$, voter~$V_i$ holds a vote~$v_i \in \{0,1\}$.\\
$Gt$ denoted the government who can identify these $n$ voters. \\
$G$ is a cyclic group of order $q$ with generator $g$.\\
$o$ is the aggregated result from $Gt$.\\
\textbf{Protocol detail in appendix~\ref{alg: homo enc based}}

\begin{theorem}\label{Security of HEV}
 If the protocol is strictly followed, $o = g^{\sum_{i=1}^{n}{v_i}} $ (Proof in appendix~\ref{Proof of HEV}).
\end{theorem}

After the threshold decryption~\ref{HEV: Threshold Decryption}, the government collects the result $o = g^{\sum_{i=1}^{n}{v_i}} $, and what we want is $\sum_{i=1}^{n}{v_i}$. As a discrete log problem, this is a $n$-bounded problem. However, the result can be verified by matching the result with $g^{0}, g^{1}, g^{2}, ..., g^{n}$. 
Generally, $n$ is bounded by the population, and the computation and comparison can be paralleled. Thus the result $\sum_{i=1}^{n}{v_i}$ can be retrieved with reasonable cost.



\section{Discussion and Improvements}\label{Improvement}

In this section we address potential issues in the design and possible solutions.

\subsection{Potential Attacks on BSV}

\subsection*{Traffic Analysis}

In BSV protocol, the voter have to publish the ballot to the block chain for it to be counted. This creates the risk of traffic analysis which an adversary can correlate the ballot with the voter by monitoring the network traffic from the voter to the blockchain. 
There are several ways to evade such attack. A Mixnet system can be used in conjunction with the blockchain. Using DC-net system can achieve similar effect but its limited through put make it less suitable for large scale voting. To further increase the anonymity set, one can designate all ballots to be casted in a certain period of time to ensure that there would be a sufficient number of voters posting their ballot at any moment.
Depending on the implementation, the voting authority $G$ can reveal the voter's identities through timing analysis if the ballots are casted immediately after the voters request signatures from $G$. Thus it is recommended to introduce a large, random delay between the signing and posting of the ballot. This can be done by making the ballot signing timeframe and voting timeframe non-overlapping.

\subsection{Potential Attacks on HEV}

The stability of homomorphic encryption voting protocol relies on the honesty of all parties involved. The design of HEV protects the privacy of voters against dishonest authority. On the other hand, the voting result can be easily invalidated by repeated voting from malicious voters. The tentative combats on these threats are explained in the following.


\subsection*{Extra Votes}
In the basic encryption setting, each individual voter is supposed to encrypt $0$ for against and $1$ for support. However, theoretically, voters are free to encrypt other numbers like $2$ or $3$ to earn extra votes. For example, with a voter group of 3, $V_1$ encrypted 3 in its voting message would dominate the aggregated result if all the other voters are honest. Since the message is encrypted by individual voters, there is no way for the government to tract this kind of cheating. 

One way to combat this is to add a zero-knowledge proof on top of the voting procedure to prove that the voter is really sending a $0$ or $1$ instead of other invalid numbers. To make the failing probability negligible, multiple tests of zero-knowledge proof are needed, in the cost of extra network traffic.

\subsection*{Cooperation Interruption}

The original HEV requires all voters to remain cooperative through the entire protocol. However, this may not hold due to various active and passive actions. For instance, voters turn offline after voting submissions and during threshold decryption (one potential reason is the network disconnetion). Moreover, some voters can decide to not send the decryption result in threshold decryption or even send something random instead of $\hat{X_i}={X_{R}}^{sK_i}$. The final result is hard to obtain in either way (or being inaccurate) because the government requires decryption results from all voters to compute $W$. We believe a stochastic approach in the protocol helps mitigate the interruption effects.

\subsubsection*{Assumption}\label{Malicious Assumption}
For simplicity, we assume that malicious voters behave like honest voters and vote against ($0$) before threshold decryption but interrupt the cooperation in threshold decryption by providing fake data (instead of using their own secret keys). We also assume that the result becomes unreliable if any malicious voters are involved in the communication. If the result we get matches the result with same group of voters and all being honest, it is considered as reliable.

\subsubsection*{Partial Public Key Generation}
A primary solution is based on partial public key generation. Specifically, the public key generated only relies on a sampled subset of the voter group, instead of the entire group as in the original HEV protocol. To achieve this, the government randomly samples $t$ public key pieces from $n$ pieces collected from $n$ voters, and generates the public key only based those $t$ pieces. Therefore, even though there are $m$ voters being interrupted, the probability of having a reliable result is $P = \frac{{{n-m}\choose t}}{{{n}\choose t}}$. For instance, when $n=50$, $m=5$ ($\frac{m}{n}=0.1$), on expectation $t=\frac{n}{2} = 25$, $P = 0.025$, which means even if malicious voters are not majority, the probability of getting a reliable result is almost $0$. 


\subsubsection*{K-Sampling}
To improve the performance of the HEV protocol with sampling, we can upgrade the primary solution. In Protocol~\ref{alg: homo enc with sanmple}, instead of sampling once and generate one public key, $Gt$ samples $k$ times and generates $k$ public keys. After the threshold decryption $Gt$ gets $k$ results, then mode is selected as the final result. We argue that if the honest voters are the majority, there are ways to get reliable result with almost $100\%$ probability by adjusting the parameter $k$.
\newline

\begin{protocol}{HEV with Sampling}
\label{alg: homo enc with sanmple} 
\textit{Additional Notations.}

$k$ is the total number times of sampling $Gt$ does, also being the number of public key pieces $pK$ uses \\
$t_j \in \{t_1, t_2,..., t_k\}$ is the number of keys chosen from voters in the $j$-th sampling (with replacement). \\
$\{ pK^{(j)}_1, pK^{(j)}_2, ..., pK^{(j)}_{t_j} \}$ is the set of chosen public key pieces in the $j$-th sampling, where $pK^{(j)}_{t_m}$, $pK^{(j)}_{t_n}$ can equal for some $m, n$.

\sbline
\textit{Procedure:}
\begin{enumerate}
  \item 
  \textbf{Public Key Generation} \text{: Same as Protocol~\ref{alg: homo enc based}}
  
  \item \textbf{Public Key Broadcast}
  \begin{enumerate}
      \item 
      Receiving all public key pieces $\{pK_{i}\}$, $i \in \{1, 2, ... n\}$.
      
      \item\label{Broadcast 1} 
      $Gt$ randomly picks $t_j$ pieces $\{ pK^{(j)}_1, pK^{(j)}_2, ..., pK^{(j)}_{t_j} \}$ where $pK^{(j)}_{t_j} \in \{pK_{i}\}$ 
      
      \item\label{Broadcast 2} 
      $Gt$ computes the $j$-th public key $pK^{(j)} = \Pi_{m=1}^{t_j}{pK^{(j)}_m}$
      
      \item 
      $Gt$ repeats \ref{Broadcast 1} - \ref{Broadcast 2} $k$ times, generating $\{ pK^{(1)}, pK^{(2)}, ..., pK^{(k)}\}$
       
      \item 
      $Gt$ broadcasts $ (pK^{(1)}, pK^{(2)}, ..., pK^{(k)}) $ to all members of $\mathcal{V}$.
  \end{enumerate}
  
  \item \textbf{Vote Delivery}
  \begin{enumerate}
      \item Same as Protocol 2. for each $pK^{(j)}, j\in\{1,2,...,k\}$, the corresponding vote for $pK^{(j)}$ is $(X_i^{(j)}, Y_i^{(j)})$ 
      \item Each $V_i$ sends all $(X_i^{(j)}, Y_i^{(j)})$ to $Gt$.
  \end{enumerate}
  
  \item \textbf{Vote Aggregation}
  \begin{enumerate}
      \item 
      $Gt$ calculates $(X_{R}^{(j)},Y_{R}^{(j)}) = (\Pi_{i=1}^{R}{X_i^{(j)}},\Pi_{i=1}^{R}{Y_i^{(j)}})$ for each pair of $(X_i^{(j)}, Y_i^{(j)})$ as in Protocol~\ref{HEV: Vote Aggregation}.
      
      \item 
      $Gt$ generates $\{(X_{R}^{(j)},Y_{R}^{(j)})\}, j \in \{1, 2, ... , k\}$.
      
  \end{enumerate}
  
  \item \textbf{Threshold Decryption}
  \begin{enumerate}
      \item 
      $Gt$ sends all $(X_{R}^{(j)},Y_{R}^{(j)}, d)$ to $\mathcal{V}$ where $d$ indicates pending decryption action from $V_{i}$.
      \item 
      Each $V_i$ sends $(\hat{X}_i^{(j)},Y_{R}^{(j)})=({{X_{R}^{(j)}}}^{sK_i}, Y_{R}^{(j)})$ to $Gt$ as in Protocol 2.\ref{HEV: Threshold Decryption}
      
      \item 
      For each $(\hat{X}_i^{(j)},Y_{R}^{(j)})$, $Gt$ computes $W^{(j)}=\Pi_{q=1}^{t_j}{\hat{X}_q^{(j)}}$, where he only multiplies the voters' feedback if $V_i$ is involved in the sampling and computation of $pK^{(j)}$, then he compute $o^{(j)} = Y_{R}/W$
      
      \item 
      $Gt$ gets $\{o^{(1)}, o^{(2)}, ..., o^{(k)} \}$, then he picks the Mode as the final result. 
      
  \end{enumerate}
  
\end{enumerate}
\end{protocol}

\begin{theorem}\label{Security of HEVS}
  If there are more (semi-)honest voters than malicious voters, the Mode of $\{o^{(1)}, o^{(2)}, ..., o^{(k)}\}$ is the reliable result. 
\end{theorem}

\begin{proof}\label{Proof of HEVS}
Instead of a rigorous proof, we present idea of the proof.

According to assumption~\ref{Malicious Assumption}, if the $j$-th sampling contains honest voters only, then the result is reliable; otherwise, if the $j$-th sampling contains any number malicious voter, we assume the result is unreliable (We ignore fake positive cases\footnote{For example, two malicious voters occasionally do the decryption for each other.}).
The idea is that, even if the same malicious voter participated in two generation steps of public keys $pK^{(e)}$ and $pK^{(f)}$ and the corresponding fake threshold decryption responses $(\hat{X}_i^{(e)},Y_{R}^{(e)})$ and $(\hat{X}_i^{(f)},Y_{R}^{(f)})$ are identical. Since $\sum{r^{(e)}}$ and $\sum{r^{(f)}}$ are different, their result $o^{(e)}$ and $o^{(f)}$ map to two different elements in the group $G$:

Suppose $(\hat{X}_i^{(e)},Y_{R}^{(e)}) = ({{X_{R}^{(e)}}}^{rd},Y_{R}^{(e)})$ and  $(\hat{X}_i^{(f)},Y_{R}^{(f)}) = ({{X_{R}^{(f)}}}^{rd},Y_{R}^{(f)})$, where $rd$ is a random number, then the two results don't equal.
\[
o^{(e)} = \frac{(g^{{sK}^{(e)}})^{\sum{r^{(e)}}} \cdot g^{\sum{v}}}{(g^{{sK}^{(e)}-rd})^{\sum{r^{(e)}}}} 
\neq 
o^{(f)} = \frac{(g^{{sK}^{(f)}})^{\sum{r^{(f)}}} \cdot g^{\sum{v}}}{(g^{{sK}^{(f)}-rd})^{\sum{r^{(f)}}}}
\]

Thus we can assume that the results of different failing sampling groups (results are unreliable) are different. If you have the same results, there is negligible probability that they are related to malicious voters. 
Thus, even two or three consistent results can be considered as the reliable result. Therefore choosing the Mode of $\{o^{(1)}, o^{(2)}, ..., o^{(k)}\}$ is a reliable result.
\end{proof}




\section{Experimental Results}\label{Experiment Results}
\begin{figure}[ht]
\centering
\includegraphics[width=0.5\linewidth]{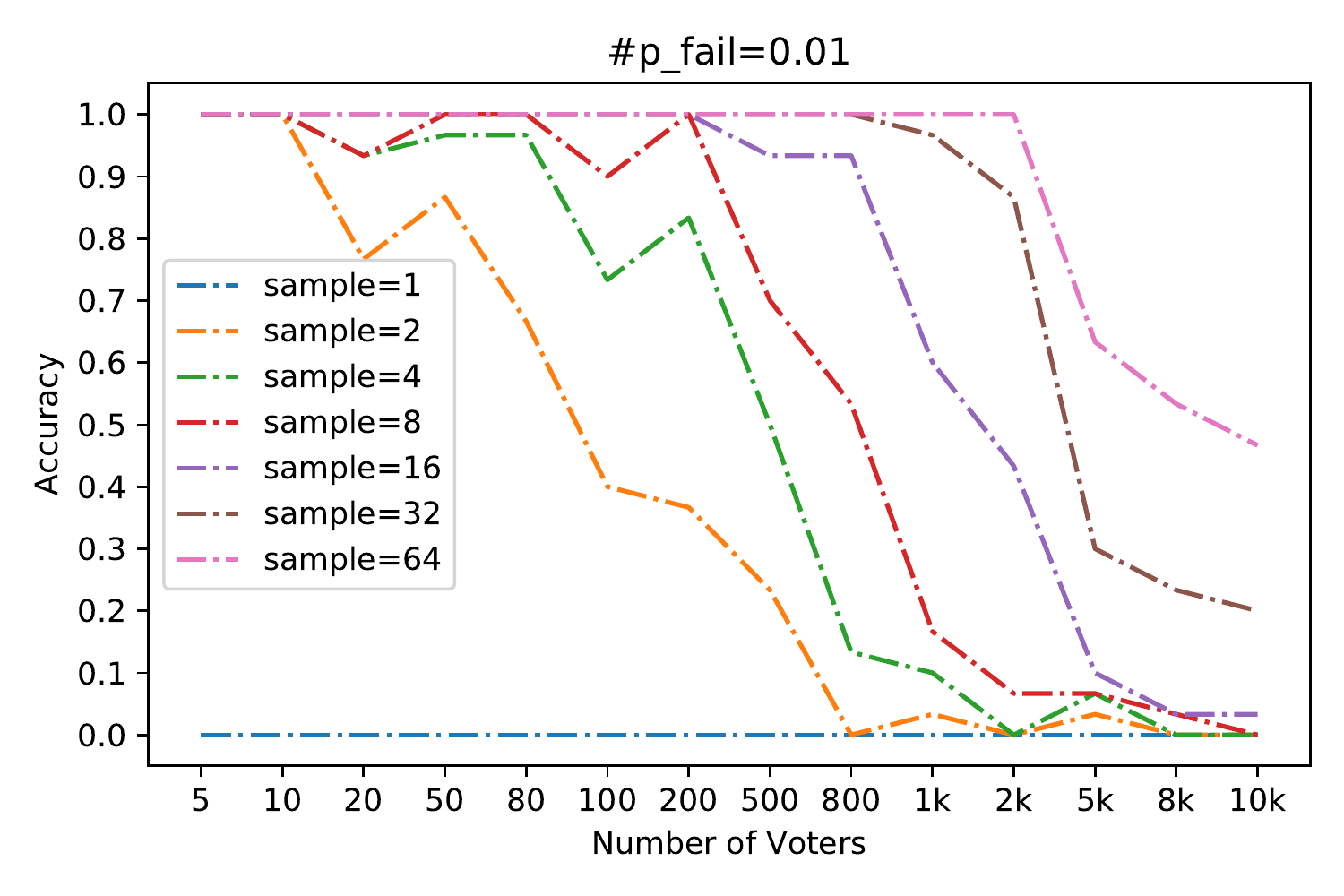}\hfill
\includegraphics[width=0.5\linewidth]{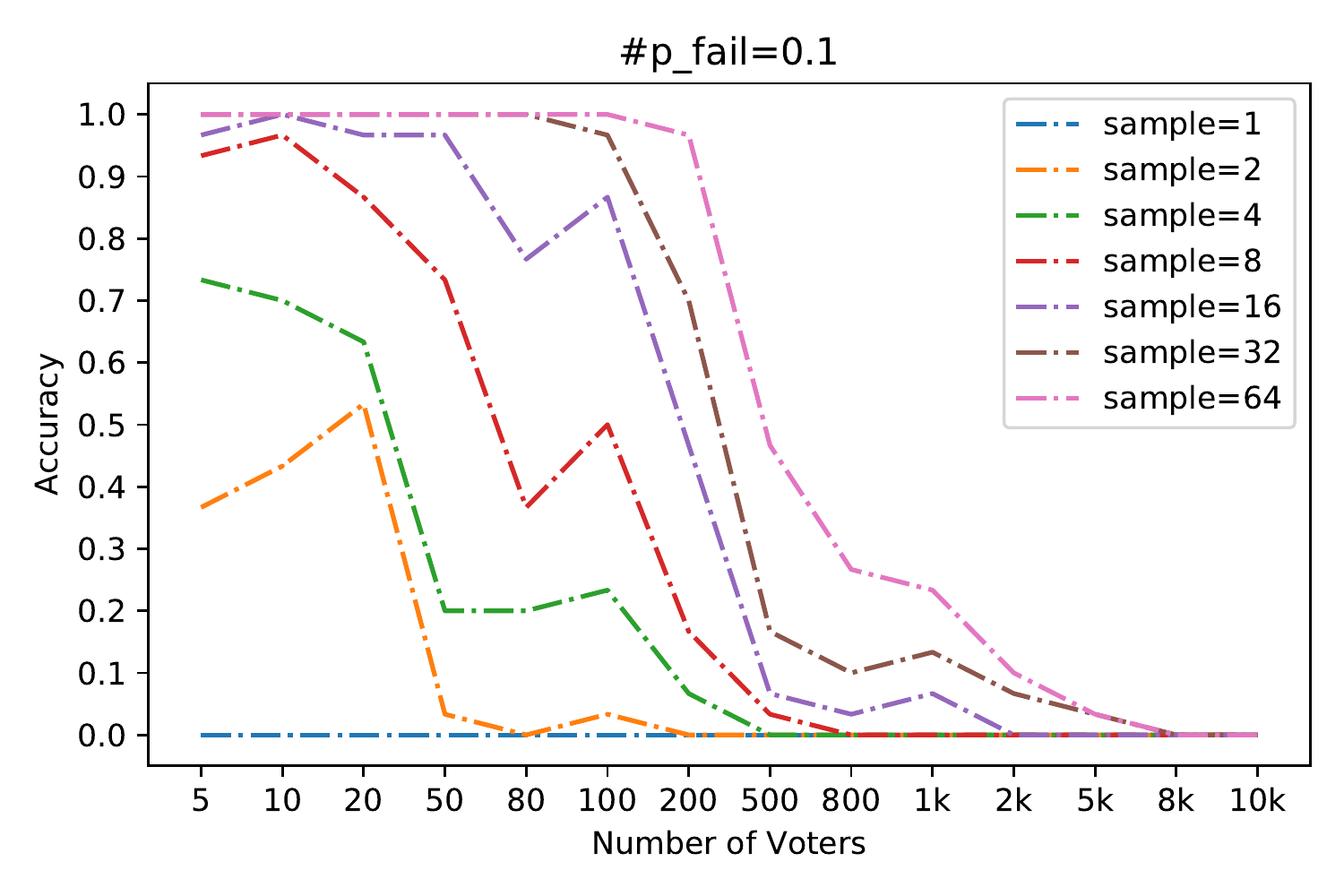}\hfill
\includegraphics[width=0.5\linewidth]{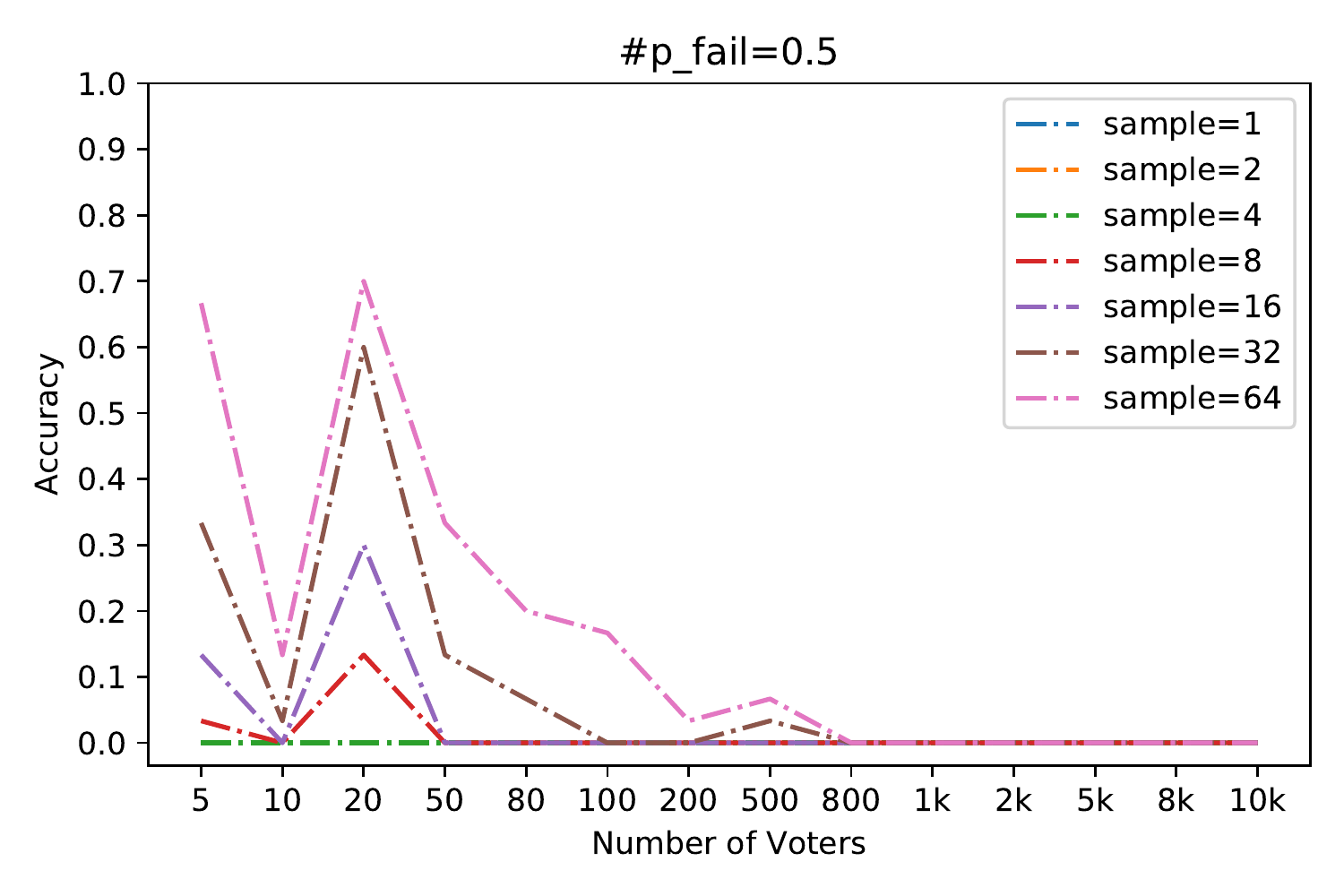}\hfill
\includegraphics[width=0.5\linewidth]{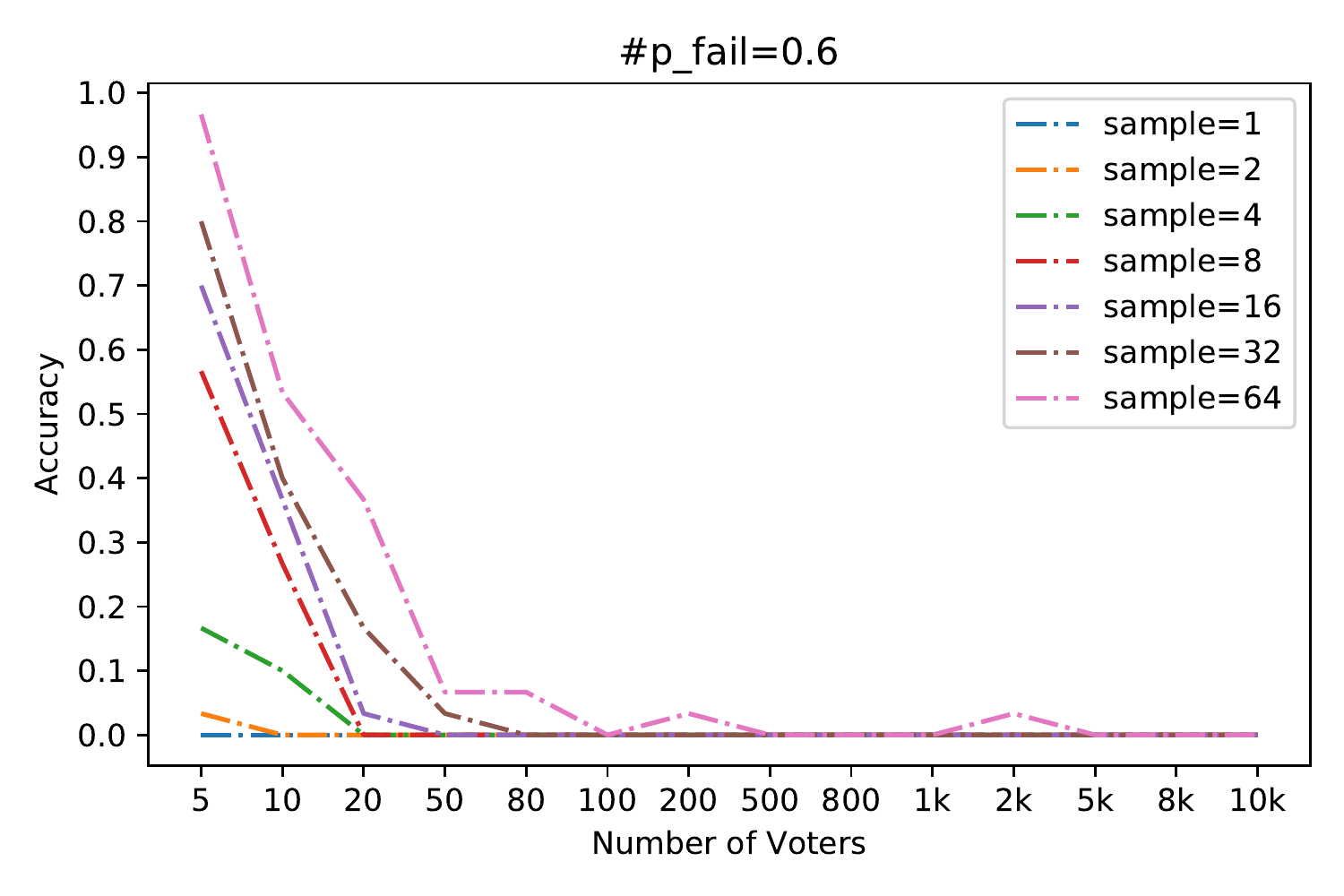}
\caption{Accuracy according to different failure ratio of the users and number of sampling times with at least 2 consistent result.}
\label{fig:experiment1}
\end{figure}

\begin{figure}[ht]
\centering
\includegraphics[width=0.5\linewidth]{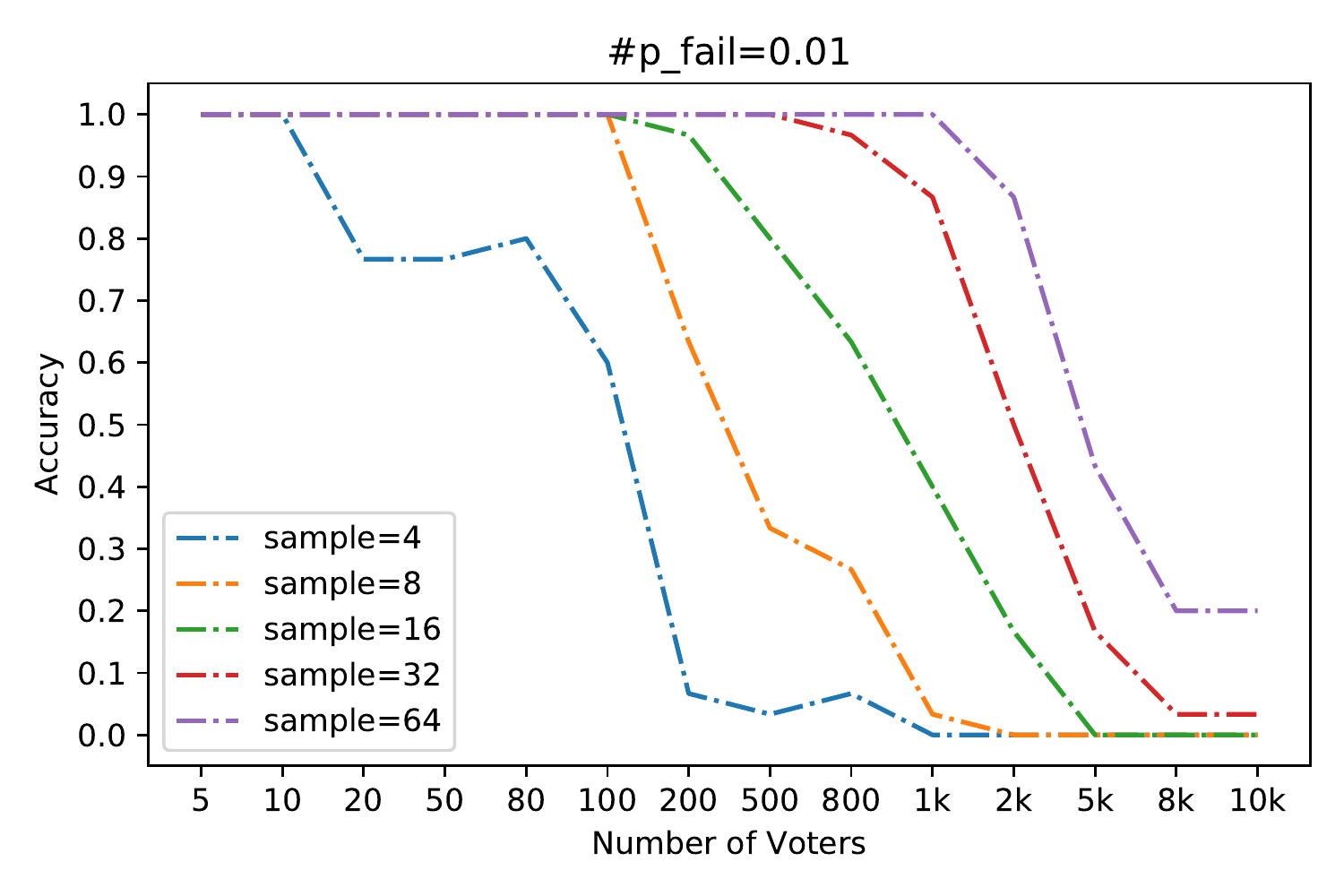}\hfill
\includegraphics[width=0.5\linewidth]{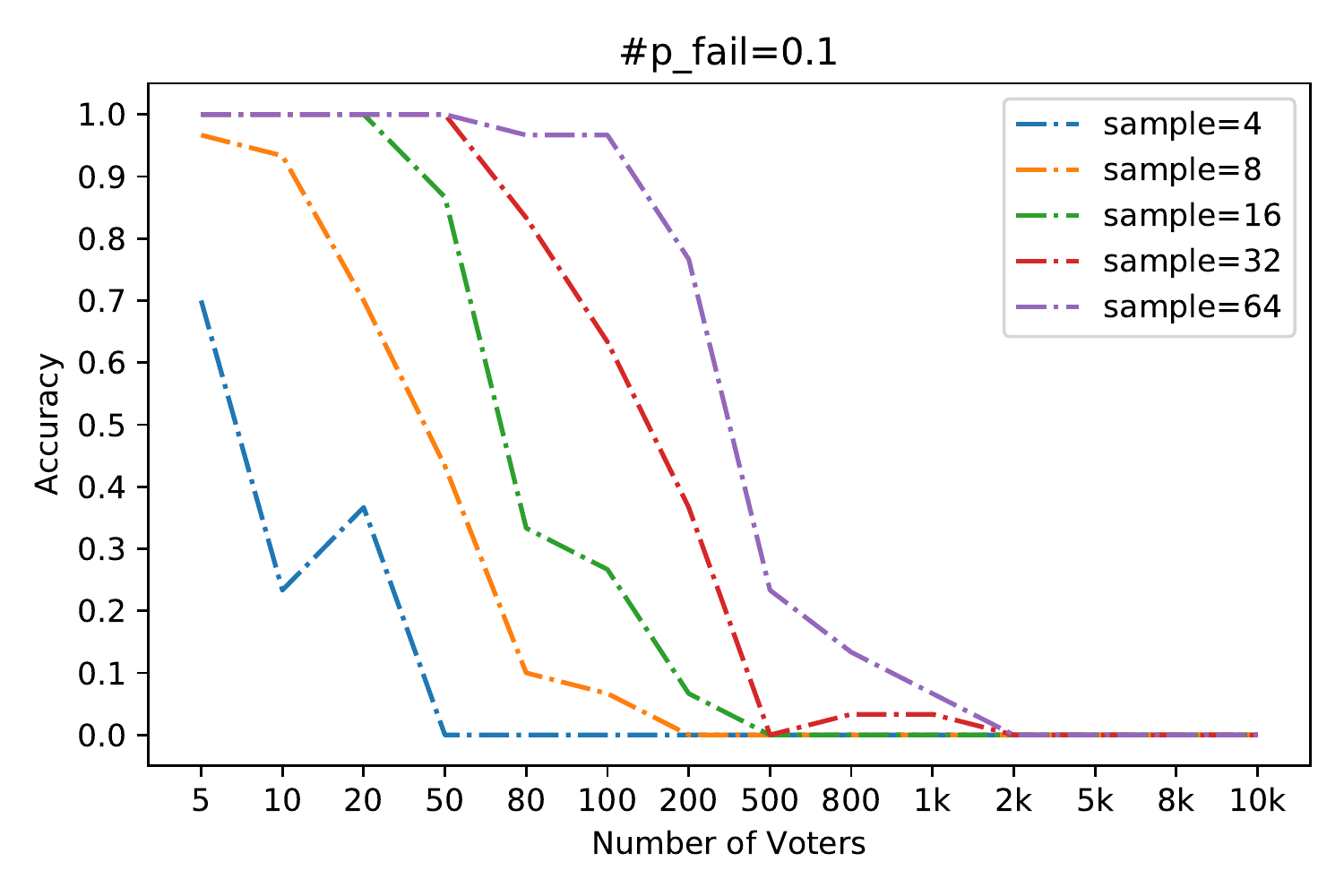}\hfill
\includegraphics[width=0.5\linewidth]{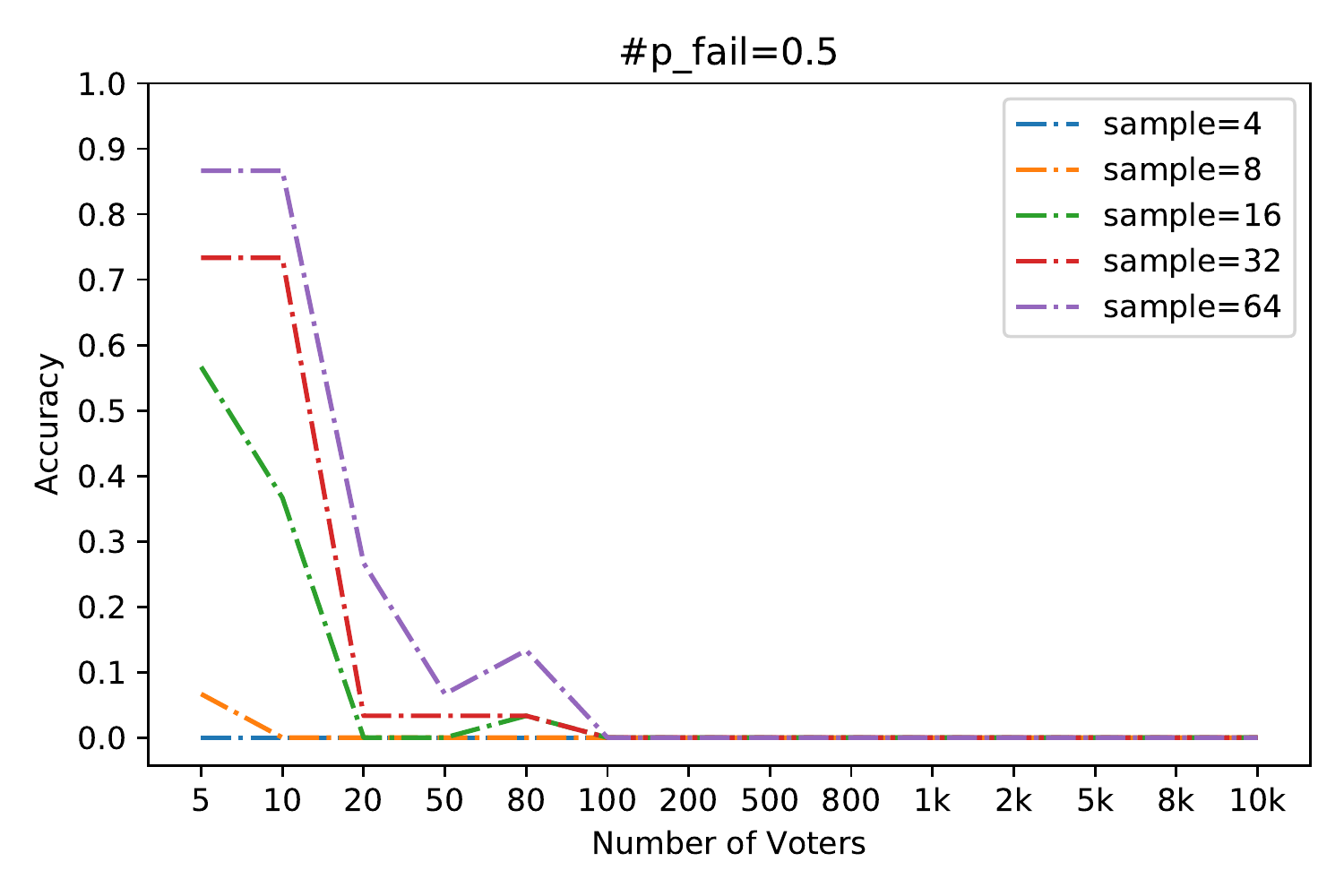}\hfill
\includegraphics[width=0.5\linewidth]{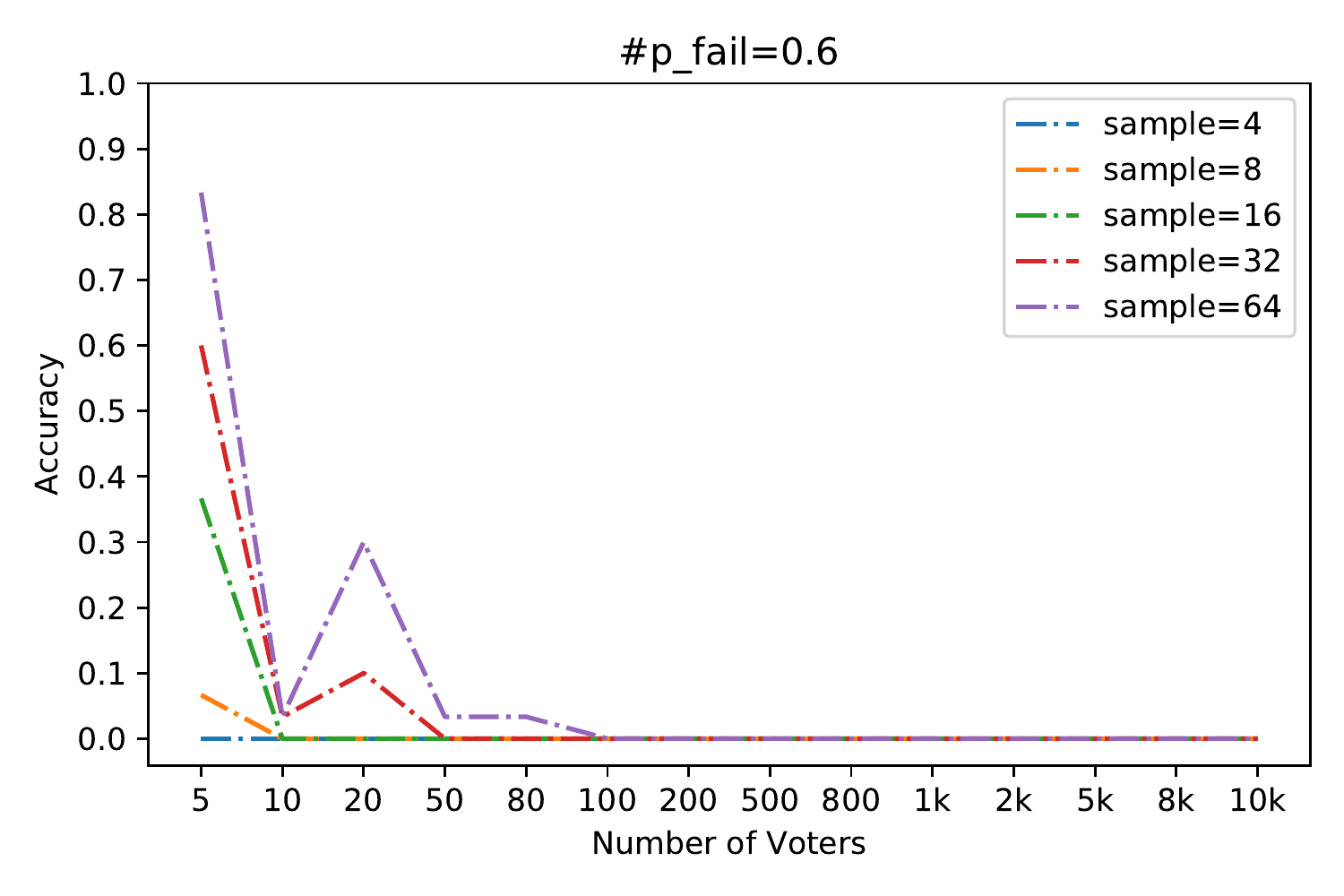}
\caption{Accuracy according to different failure ratio of the users and number of sampling times with at least 3 consistent result.}
\label{fig:experiment_3samples}
\end{figure}

According to our assumption~\ref{Malicious Assumption} and Theorem \ref{Security of HEVS}, we simulate experiments based on the number of voters $n$, the probability that a voter is malicious $p_{fail}$, and the number of sampling $k$. Figure~\ref{fig:experiment1} shows the performance of getting at least two consistent results and Figure~\ref{fig:experiment_3samples} shows the performance of getting at least three consistent results. The result is averaged from $3$ random seeds. Our protocol achieves almost $100\%$ accuracy for no more than $500$ voters when $p_{fail} = 0.01$ when sampling more than $6$ times. The protocol is almost accurate for no more than $200$ voters when $p_{fail} = 0.1$ if sampling more than $16$ times, which improves the performance of almost zero for the primary solution instance. For more than half malicious voters, the protocol doesn't remain effectiveness.



\section{Conclusion}
This work provides two approaches for privacy-preserving electronic voting system. Both of them are privacy preserving under semi-honest government. The BSV protocol widely accepted in current literature, which is stable under semi-honest and malicious voters. The HEV protocol is stable under semi-honest voter and the updated HEVS provides stability under malicious voters. There is no solution for a malicious government who rejects all communications.


\bibliography{mybib}
\bibliographystyle{plain}

\appendix

\section{Protocol BSV}
\begin{protocol}{Blind Signature Voting (BSV)}
\label{alg: blind sign based} 
\textit{Notation.} 
$\mathcal{V} = \{V_1, V_2, ... V_i, ... V_{n}\}$ is the group of eligible voters. $\forall i \in \{1, 2, ... n\}$, voter~$V_i$ holds a vote~$m_i$.

$Gt$ denoted the government who can identify these $n$ voters. \\
$B$ denoted a decentralized blockchain

\sbline
\textit{Objective.} Parties jointly compute $\sum_{i=1}^n{v_i}$ without revealing any $v_i$ .
\sbline
\textit{Procedure:}
\begin{enumerate}
  
  
  \item \textbf{Public Key Generation}
  \begin{enumerate}
    \item
    $Gt$ generates a public key $pK$ and private key $sK$.

    \item
    $Gt$ announces $pK$ to the public.
  \end{enumerate}
  
  \item \textbf{Ballot Blinding}
  \begin{enumerate}
      \item 
       Each voter generates a ballot in the form of $(m_i, r)$, where $m_i$ is the ballot content and $r$ is a large randomly generated number. 
      \item
      Through a secret blinding factor $b$, each voter blinds the ballot into $(m_i, r)'$ and sends it to $Gt$.
  \end{enumerate}
  
  \item \textbf{Signature}
  \begin{enumerate}
      \item 
      After verifying the voter's eligibility, $G$ produces signatures $s'$ on the blinded ballots $(m_i, r)'$ and sends $s'$ back to the voters.
  \end{enumerate}
  
  \item \textbf{Ballot Unblinding}
  \begin{enumerate}
      \item 
      Using the blinding factor $b$, each voter unblinds $s'$ into $s$, where $s$ equals $sig((m,r))$.
      \item\label{BSV: send to blockchain}
      The voters send the signed ballot $(m, r, s)$ to the blockchain $B$.
  \end{enumerate}
  
  \item \textbf{Ballot Verification}
  \begin{enumerate}
    \item
    $B$ accept a given ballot $(m, r, s)$ if and only if:
      
      The signature $s$ can be verified with $pK$.
      
      The random number $r$ in $(m, r, s)$ did not appear in any of the previous ballots.
  \end{enumerate}
  
\end{enumerate}
\end{protocol}

\section{Protocol HEV}
\begin{protocol}{Homomorphic Encryption Voting (HEV)}
\label{alg: homo enc based} 
\textit{Notation.} $\mathcal{V} = \{V_1, V_2, ... V_i, ... V_{n}\}$ is the group of eligible voters. $\forall i \in \{1, 2, ... n\}$, voter~$V_i$ holds a vote~$v_i \in \{0,1\}$.

$Gt$ denoted the government who can identify these $n$ voters. \\
$G$ is a cyclic group of order $q$ with generator $g$.\\
$o$ is the aggregated result from $Gt$.\\
\sbline
\textit{Objective.} Parties jointly compute $\sum_{i=1}^n{v_i}$ without revealing any $v_i$ .
\sbline
\textit{Procedure:}
\begin{enumerate}
  
  \item \label{HEV: Public Key Generation} \textbf{Public Key Generation}
  \begin{enumerate}
    \item
    Each voter $V_i$ chooses a random number $sK_i$ from $\{1,2,...,q-1\}$.

    \item
    Each voter $V_i$ computes public key piece $pK_i = g^{sK_i}$ and send it to $G$.

  \end{enumerate}
  
  \item \textbf{Public Key Broadcast}
  \begin{enumerate}
      \item 
      After receiving all the $pK_i$ ($i \in \{1, 2, ... n\}$), $Gt$ computes the public key $pK = \Pi_{i=1}^{n}{pK_i}$  
      \item
      $Gt$ broadcasts $pK$ to all members of $\mathcal{V}$.
  \end{enumerate}
  
  \item \textbf{Vote Delivery}
  \begin{enumerate}
      \item 
      After receiving $pK$, each voter $V_i$ generates a random number $r_i$ from $G$, and computes $X_i = g^{r_i}$,  $Y_i = pK^{r_i} \cdot g^{v_i}$.
      
      \item
      Each $V_i$ sends $(X_i, Y_i)$ to $Gt$.
  \end{enumerate}
  
  \item\label{HEV: Vote Aggregation} \textbf{Vote Aggregation}
  \begin{enumerate}
      \item 
      $Gt$ calculates $(X_{R},Y_{R}) = (\Pi_{i=1}^{R}{X_i},\Pi_{i=1}^{R}{Y_i})$. with $R$ is the number of votes received. the computing stops when $R$ equals to $n$.
  \end{enumerate}
  
  \item\label{HEV: Threshold Decryption} \textbf{Threshold Decryption}
  \begin{enumerate}
      \item 
      $Gt$ sends $(X_{R},Y_{R},d)$ to $\mathcal{V}$ where $d$ indicates pending decryption action from $V_{i}$.
      \item
      Each $V_i$ sends $(\hat{X_i},Y_{R})=({X_{R}}^{sK_i}, Y_{R})$ to $Gt$ \newline
      ($V_i$ can first check whether the received $(X_R,Y_R)$ equals$(X_i,Y_i)$, if so, he refuses to decrypt.)
      
      \item
      After receiving all the $(\hat{X_i}, Y_{R})$, $Gt$ computes $W=\Pi_{i=1}^{n}{\hat{X_i}}$, then he compute $o = Y_{R}/W$
      
  \end{enumerate}
  
\end{enumerate}
\end{protocol}

\section{Proof of Theorem~\ref{Security of HEV}}
\begin{proof}
\begin{align*}
    W = \Pi_{i=1}^{n}{\hat{X_i}} &= \Pi_{i=1}^{n}{{X_{R}}^{sK_i}} \\
    &= X_{R}^{\sum_{i=1}^{n}{sK_i}} \\
    &= (\Pi_{i=1}^{R}{X_i})^{\sum_{i=1}^{n}{sK_i}} \\
    &= (\Pi_{i=1}^{R}{g^{r_i}})^{\sum_{i=1}^{n}{sK_i}} \\
    &= (g^{\sum_{i=1}^{R}{r_{i}}})^{\sum_{i=1}^{n}{sK_i}} 
\end{align*}
 
\begin{align*}
    Y_{R} &= \Pi_{i=1}^{R}{Y_i} \\
    &= \Pi_{i=1}^{R}{pK^{r_i}\cdot g^{v_i}} \\
    &= (pK^{\sum_{i=1}^{n}{r_i}}) \cdot g^{\sum_{i=1}^{R}{v_i}} \\
    &=  [(\Pi_{i=1}^{n}{pK_i})^{\sum_{i=1}^{n}{r_i}}] \cdot g^{\sum_{i=1}^{R}{v_i}} \\
    &= [(\Pi_{i=1}^{n}{g^{sK_i}})^{\sum_{i=1}^{n}{r_i}}] \cdot g^{\sum_{i=1}^{R}{v_i}} \\
    &= [(g^{\sum_{i=1}^{n}{sK_i}})^{\sum_{i=1}^{n}{r_i}}] \cdot g^{\sum_{i=1}^{R}{v_i}} \\
    &= W \cdot g^{\sum_{i=1}^{R}{v_i}}
\end{align*}

 Then $o =Y_{R}/W =g^{\sum_{i=1}^{n}{v_i}} $ if $R=n$

\end{proof}\label{Proof of HEV}

\section{Experiments}

\end{document}